\documentclass[]{spie}  

\usepackage{amsmath,amsfonts,amssymb}
\usepackage{graphicx}
\usepackage[colorlinks=true, allcolors=blue]{hyperref}

\title{Dissociative photoionization of EUV lithography photoresist models}

\author[a]{Marziogiuseppe Gentile}
\author[b]{Marius Gerlach}
\author[c]{Robert Richter}
\author[a]{Michiel J. van Setten}
\author[a]{John S. Petersen}
\author[a]{Paul van der Heide}
\author[a]{Fabian Holzmeier}

\affil[a]{imec, Kapeldreef 75, 3000 Leuven, Belgium}
\affil[b]{Institute for Physical and Theoretical Chemistry, University of W{\"u}rzburg, Am Hubland, 97074 W{\"u}rzburg, Germany}
\affil[c]{Elettra Sincrotrone Trieste, 500 in Area Science Park, 314149 Basovizza, Italy}

\authorinfo{Further author information: (Send correspondence to F.H.)\\F.H.: E-mail: fabian.holzmeier@imec.be}

\begin{document} 
\maketitle

\begin{abstract}
The dissociative photoionization of \textit{tert}-butyl methyl methacrylate, a monomer unit found in many ESCAP resists, was investigated in a gas phase photoelectron photoion coincidence experiment employing extreme ultraviolet (EUV) synchrotron radiation at 13.5 nm. It was found that the interaction of EUV photons with the molecules leads almost exclusively to dissociation. However, the ionization can also directly deprotect the ester function, thus inducing the solubility switch wanted in a resist film. These results serve as a building block to reconstruct the full picture of the mechanism in widely used chemically amplified resist thin films, provide a knob to tailor more performant resist materials, and will aid interpreting advanced ultrafast time-resolved experiments.
\end{abstract}


\section{INTRODUCTION}
\label{sec:intro}  

The move from deep ultraviolet (DUV) photolithography using 248-193~nm (4.8-6.4~eV) to extreme ultraviolet (EUV) at 13.5~nm (92~eV) means that the way light interacts with a photoresist thin film has changed fundamentally. While DUV light selectively activates chemical bonds in the resist material by resonant excitation, the high photon energy of EUV intrinsically triggers ionization events, but this process only has a low local selectivity. The primary photoionization events furthermore lead to a complex radiation chemistry in a resist thin film. In order to design potent resist materials for EUV lithography that are suited for imaging feature sizes below 20~nm, it is crucial to understand and ultimately control the physical and chemical processes in a photoresist film imaged with EUV radiation. In this paper, the dissociative photoionization of \textit{tert}-butyl methylacrylate (TBMA), a monomer unit that is widely used in (co-)polymers of chemically-amplified resists (CARs), is investigated using photoelectron photoion coincidence (PEPICO) spectroscopy in the gas phase. This reduces the complexity of the chemistry by focusing only on the initial step in the interaction of EUV photons with resists and gaining deep fundamental insights that would not be accessible without this isolated view. Together with further complementary experiments, these insights are one basic building block in deciphering the full chemical and physical processes in EUV photolithography.

The interaction of a EUV photon with a photoresist material leads to photoionization, i.e., the emission of a primary photoelectron. The ionization energy of organic molecules lies typically in the 10~eV range. Thus, the produced photoelectrons can possess a significant amount of kinetic energy at 92~eV photon energy, which leads to a second important reaction step in the EUV exposure mechanism: The primary electrons interact further with the material inducing a chain reaction of more ionization events and the creation of secondary electrons albeit with less kinetic energy. These low energy secondary electrons are assumed to mainly drive the radiation chemistry in EUV resists by electron impact dissociation and electron attachment processes. In CARs, either of these processes can lead to the generation of the photoacid, which then catalyses the deprotection of a functional group of the base polymer and therefore leads to a solubility switch. Here, also diffusion processes become relevant that make it even more challenging to obtain the desired resolution of the imaged pattern. Most of the studies on EUV photochemistry follow the pathway of the electrons generated by photoionization. However, removal of an electron can weaken a chemical bond and it is possible that a fragmentation is directly induced at this step leading to potentially reactive dissociation products. Although the electron chemistry is probably the main driving force in the solubility switch, the unavoidable fragmentations by photoionization must not be neglected since they will often lead to unwanted side reactions deteriorating the performance of a resist. 

The complex reaction sequences in the condensed phase make it very challenging to investigate EUV photochemistry directly in thin resist films with established analysis methods.\cite{Kozawa2010} An alternative approach is therefore to use complementary experimental techniques and investigate the various processes individually in an isolated gas phase environment first in order to get fundamental insights into the chemistry occurring in each step.\cite{Grzeskowiak2017,Kostko2018} Gas phase photoelectron spectroscopy and photoionization mass spectrometry yield information on the initial photoionization step,\cite{Kostko2018,Wu2019,Thakur2020} whereas the chemistry induced by secondary low energy electrons can be investigated using focused electron beams to study dissociative electron impact ionization and dissociative electron attachment in the gas phase.\cite{Rathore2021} The deep insights gained in such experiments can then be used to understand experiments on thin films, such as photoemission, which also adds information on the kinetic energy distribution of secondary electrons, \cite{Ma2019,Zhang2021,Kostko2022}  or outgassing experiments. The latter can identify charged and neutral fragment species that leave the surface of the thin film during EUV exposure.\cite{Pollentier2018} Techniques like infrared and x-ray photoemission spectroscopy
before and after exposure of the film provide information on the chemical net change in the resist upon exposure. Combining this uncomplete list of techniques helps to draw a more complete picture of the photoresist chemistry in EUV lithography.  

In this paper, we report on PEPICO experiments of a prototypical CAR monomer, \textit{tert}-butyl methacrylate, in the gas phase providing novel fundamental insights that can be transferred to better understand the photoionization of resists by EUV photons. PEPICO combines photoelectron spectroscopy and photoionization mass spectrometry by detecting electrons and ions in coincidence.\cite{Baer2000} Hereby, photoelectron spectra yielding the kinetic energy distribution of the primary photoelectrons are obtained as well as mass spectra, from which fragmentation products and their respective ratio produced by dissociative photoionization are obtained. In addition, PEPICO allows to combine these two datasets by correlating electrons and ions for each single ionization event. In practice, the intensity of the photon source should be so low that one photon pulse does not lead to more than one ionization event in order to guarantee an unambiguous assignment of electrons and ions. Consequently, PEPICO experiments are best conducted with light sources which operate at a high repetition rate, such as synchrotron storage rings with repetition rates in the MHz range, in order to balance acquisition time and statistics. The appearance of fragment ions can then be assigned to electrons with a certain kinetic energy in the photoelectron spectrum. Through PEPICO it can be understood which electronic states lead to which fragment ions. For photoresists there might be fragmentation channels, which actively contribute to the solubility switch by dissociative photoionization, while others lead to reactive side products. PEPICO experiments can give an idea on how to try to change the electronic structure of the resist/molecule, e.g., by different functional groups, in order to enhance (suppress) the formation of wanted (unwanted) dissociation products. 

TBMA is a monomer used in many common CARs and containing a \textit{tert}-butyl ester side group, which is hydrolyzed upon exposure with EUV (and DUV) light leading to a higher solubility in protic solvents. Here we find that TBMA almost exclusively undergoes dissociative photoionization at 92~eV and induces a number of fragmentations, which could decrease the performance of resists containing TBMA monomers. However, one of the channels yields directly the deprotected free acid in a McLafferty rearrangement and the PEPICO technique combined with quantum chemical computations allowed to assign this dissociation channel mainly to ionization out of the HOMO-2 orbital.

\section{EXPERIMENTAL AND THEORETICAL DETAILS}
\label{sec:exp}  

The experiments were conducted at the low energy branchline of the Gas Phase beamline\cite{Prince1998} of the Elettra storage ring (Trieste, Italy). The beamline disposes over a 4.5~m long undulator with a period of 12.5~cm delivering linearly polarized light tunable between 13 and 1000 eV. A toroidal prefocusing mirror focuses the light onto the entrance slit of a variable angle spherical grating monochromator featuring five gratings to cover the beamline’s full photon energy range at a resolving power of E/$\Delta$E $\approx$ 10$^4$. For the experiments presented here conducted at 92 eV photon energy a gold coated 1200~mm$^{-1}$ grating was used. After the exit slit of the monochromator the beam is deflected into the branch line by a plane mirror and refocused using a toroidal mirror. The angles of the two mirrors limit the maximum photon energy available at the branchline to $\approx$ 250~eV. The spot in the interaction region between the repeller and extractor electrodes of the PEPICO spectrometer\cite{Plekan2008} is elliptical ($\approx$~400x250~µm$^2$). 

\begin{figure} [ht]
\begin{center}
\begin{tabular}{c} 
\includegraphics[width=16cm]{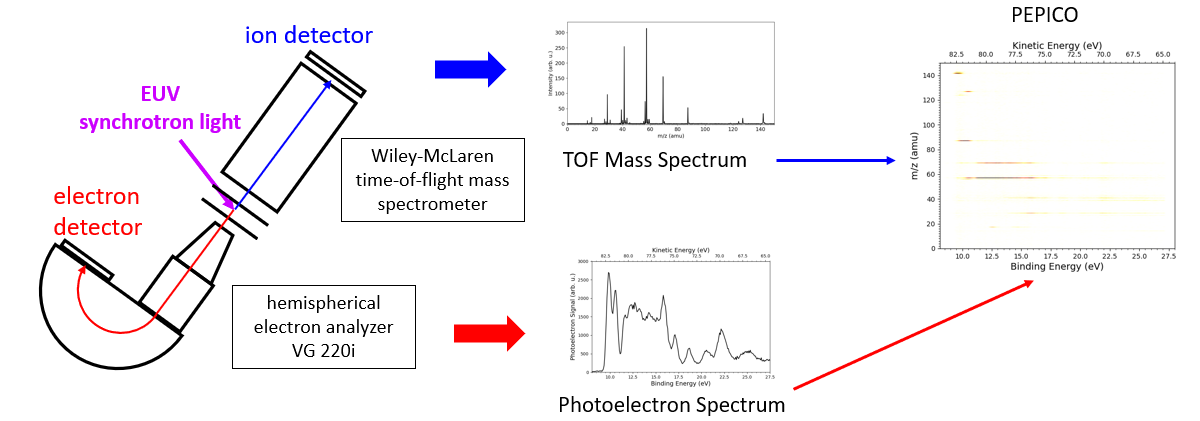}
\end{tabular}
\end{center}
\caption{\label{fig:experiment} Experimental setup of the PEPICO experiment at the Elettra Gas Phase beamline. The molecules interact with EUV synchrotron radiation in the center and electrons and ions are detected in a hemispherical electron analyzer and time-of-flight mass spectrometer, respectively. The experiment yields independent ion mass and photoelectron spectra, as well as PEPICO data combining the information of electrons and ions originating from the same ionization events. }
\end{figure}

Electrons pass through a 90~\% transmission gold mesh and are detected continuously in a 150~mm mean radius hemispherical electron energy analyzer (VG 220i) mounted behind the repeller of the time-of-flight spectrometer (TOF). The analyzer is equipped with a 2D crossed delay line detector.\cite{Cautero2008} Ions are detected in a home-made Wiley-McLaren TOF spectrometer on the opposite side of the electron analyzer. A schematic of the PEPICO setup is shown in \ref{fig:experiment}. 

The electron and ion mass spectrometers can also be operated independently for recording photoelectron spectra and time-of-flight mass spectra, respectively. Photoelectron spectra are recorded by scanning the kinetic energy for the electrons passing the analyzer in steps at a selected pass energy. In this experiment pass energies of 5~eV and 20~eV were used for the low and high binding energy regions, respectively, resulting in spectrometer energy bandwidths of 100 (400)~meV. In both cases the signal was averaged over 3~s per step and integrated over the 2D detector of the electron analyzer. In case of independent operation of the TOF spectrometer the extraction field pulses are triggered by a 1~kHz pulse generator (Stanford Research DG535). The repeller and extractor electrodes of the mass spectrometer are driven by an external trigger and polarized with antisymmetric voltages (manufacturer Directed Energy Inc., model PVM4210) producing typically an extraction field of 700~V/cm. Flight times are recorded using a multichannel time-to-digital converter (TDC, model AM-GPX, ACAM Messelectronic).

In a PEPICO experiment, the amplified and discriminated signal of the electron analyzer provides the trigger for the ion extraction pulse. False coincidences are estimated by randomly pulsing the extraction field at 100~Hz and are counted in a separate channel of the TDC. An ion signal then serves as the stop signal. Since the trigger can have two different sources, the two trigger signals are combined in an ”or” logic unit before the TDC. The measured spectra are normalized to the number of starts provided by the photoelectrons and the false coincidences are normalized to the number of random starts. The normalized spectra are then subtracted from each other to obtain the final PEPICO data.\cite{Pruemper2007} In coincidence detection mode the pass energy was set to 20~eV and coincidence spectra were recorded for central kinetic energies in steps of 1.6~eV. The 2D electron images at every step were then put together to obtain the complete PEPICO data. The final data shown here were collected by acquiring over 300~min per kinetic energy at an ion count rate of $\approx$ 28~kcts/s. The data analysis was performed using a home-written Python code.\cite{PEPICO_Elettra_VG}
Onset energies for the various fragmentation channels were determined from a linear fit of the rising part of the respective fractional abundance curve as a function of the binding energy.

\textit{Tert}-Butyl methacrylate was purchased from Sigma-Aldrich (98~\% purity) and used without any further purification. The liquid sample was mounted in an external sample reservoir and held at a constant temperature of 8$^\circ$C. An effusive beam of the sample vapor was introduced through a stainless steel needle placed between the extractor and repeller 2~mm from the photon beam. The background pressure in the experimental chamber is typically $\cdot$10$^{-7}$~mbar and the pressure was set to 3$\cdot$10$^{-6}$~mbar during the experiment. 

Molecular geometries of the neutral and cationic parent molecule, fragments, and transition states were optimized with density functional theory (DFT) on the PBE0/def2-TZVP level of theory using Turbomole 7.2.\cite{Turbomole} Transition states were located by relaxed coordinate scans of certain bond lengths and angles. All energies given in this article are referenced to the electronic ground state energy of TBMA. Orbital binding energies were calculated using single-shot $GW (G_0W_0)$ starting from Kohn-Sham orbitals and eigenvalues calculated by means of self-consistent field DFT.\cite{Hedin1965,Setten2013}

\section{RESULTS}
\label{sec:results}

\begin{figure} [ht]
\begin{center}
\begin{tabular}{c} 
\includegraphics[width=15cm]{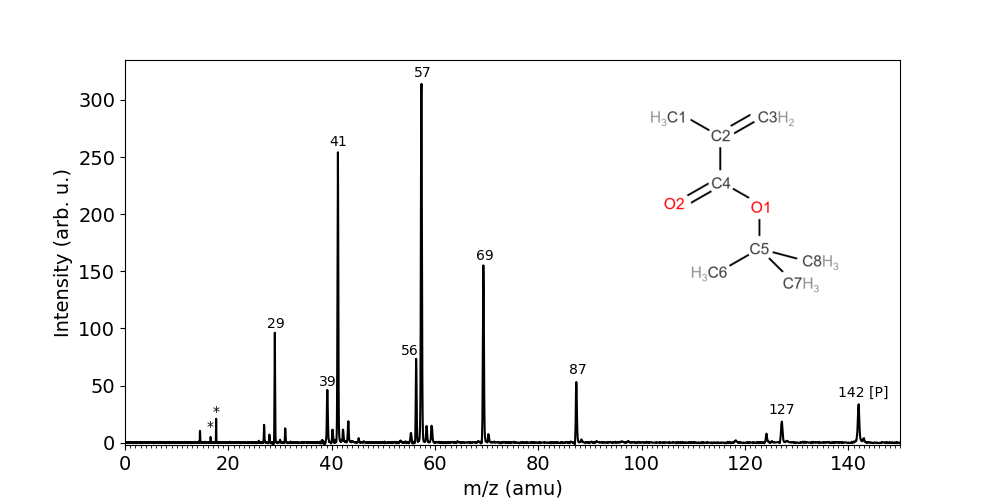}
\end{tabular}
\end{center}
\caption{\label{fig:tofms} Time-of-flight mass spectrom of \textit{tert}-butyl methacrylate recorded at 92~eV photon energy. The peaks at m/z=17 and 18~amu marked with an asterisk are assigned to a contamination by water.} 

\end{figure}

The time-of-flight mass spectrum depicted in Fig. \ref{fig:tofms} was recorded at a photon energy of 92~eV and using the random trigger mode of the mass spectrometer. It clearly shows that dissociative photoionization is dominating in the EUV and only about 3~\% of the parent ions (m/z=142~amu) do not undergo fragmentation. The three most abundant daughter ions at m/z=57~amu, m/z=41~amu, and m/z =69~amu together make up more than 60~\% of the spectrum and their origin can be explained by simple bond cleavage of O1-C5, C2-C4, and C4-O1, respectively (see the structure in Fig. \ref{fig:tofms} for the atom labeling).

Seven further ions show an important fractional abundance of 2 \% or higher, but not all of them can be explained by a simple scission of a single bond. Table \ref{tab:table_branching} lists the fractional abundance of the ten most intense fragment ions alongside the sum formula of the ionic fragment and possible neutral fragments. The fractional abundance was calculated as the integral of the corresponding peak in the mass spectrum divided by the integral of the complete spectrum excluding the peaks at m/z=17, and 18~amu, which originate from a water contamination (H$_2$O$^+$ and OH$^+$, respectively).

\begin{table} [ht]
\caption{\label{tab:table_branching}Branching ratios and chemical formulas of the ionic and neutral fragments for dissociative photoionization at 92 eV.}
\begin{center}
\begin{tabular}{|r|c|c|c|r|}
\hline
\rule[-1ex]{0pt}{3.5ex} m/z & Ionic Fragment & Neutral Fragment & Abundance (\%) & Onset Energy (eV) \\
\hline
\rule[-1ex]{0pt}{3.5ex} 142 & C$_8$H$_{14}$O$_2\!^{\cdot+}$ (P) & - & 3.4 & 9.2$\pm$0.1\\
\hline
\rule[-1ex]{0pt}{3.5ex} 127 & C$_7$H$_{11}$O$_2\!^+$ & CH$_3\!^\cdot$ & 1.9 & 9.9$\pm$0.2\\
\hline
\rule[-1ex]{0pt}{3.5ex} 87 & C$_4$H$_7$O$_2\!^+$ & C$_4$H$_7\!^\cdot$ & 4.1 & 9.4$\pm$0.2\\
\hline
\rule[-1ex]{0pt}{3.5ex} 69 & C$_4$H$_5$O$\!^+$ & C$_4$H$_9$O$\!^\cdot$ & 13.2 & 10.4$\pm$0.5 \\
\hline
\rule[-1ex]{0pt}{3.5ex} 57 & C$_4$H$_9\!^+$ & C$_4$H$_5$O$_2\!^\cdot$ & 27.4 & 9.8$\pm$0.5 \\
\hline
\rule[-1ex]{0pt}{3.5ex} 56 & C$_4$H$_8\!^{\cdot+}$ & C$_4$H$_6$O$_2$ or C$_4$H$_5$O$_2\!^\cdot$+H$^\cdot$ & 5.4 & 9.6$\pm$0.2\\
\hline
\rule[-1ex]{0pt}{3.5ex} 41 & C$_3$H$_5\!^+$ & "C$_5$H$_9$O$_2\!^\cdot$" & 21.2 & 13.1$\pm$1.0\\
\hline
\rule[-1ex]{0pt}{3.5ex} 39 & C$_3$H$_3\!^+$ & "C$_5$H$_{11}$O$_2\!^\cdot$" & 4.7 & 16.3$\pm$1.0 \\
\hline
\rule[-1ex]{0pt}{3.5ex} 29 & C$_2$H$_5\!^+$ & "C$_6$H$_{12}$O$_2\!^\cdot$" & 6.6 & 13.4$\pm$1.0\\
\hline
\end{tabular}
\end{center}
\end{table}
From a lithography perspective the occurrence of the fragment ion m/z=56~amu is particularly interesting, since the corresponding neutral fragment could be methacrylic acid C$_3$H$_5$COOH, i.e., formed through deprotection of the ester side group, which leads to the solubility switch in a chemically-amplified photoresist containing TBMA as a monomer unit. However, the fragment ion m/z=56~amu can also originate from a sequential mechanism, in which m/z=57~amu, the most abundant daughter ion in the photoionization of TBMA, undergoes a hydrogen atom loss. In the following we will focus on those two daughter ions only. An extended discussion of all further important fragmentation channels will be covered in a forthcoming publication.

\begin{figure} [ht]
\begin{center}
\begin{tabular}{c} 
\includegraphics[width=15cm]{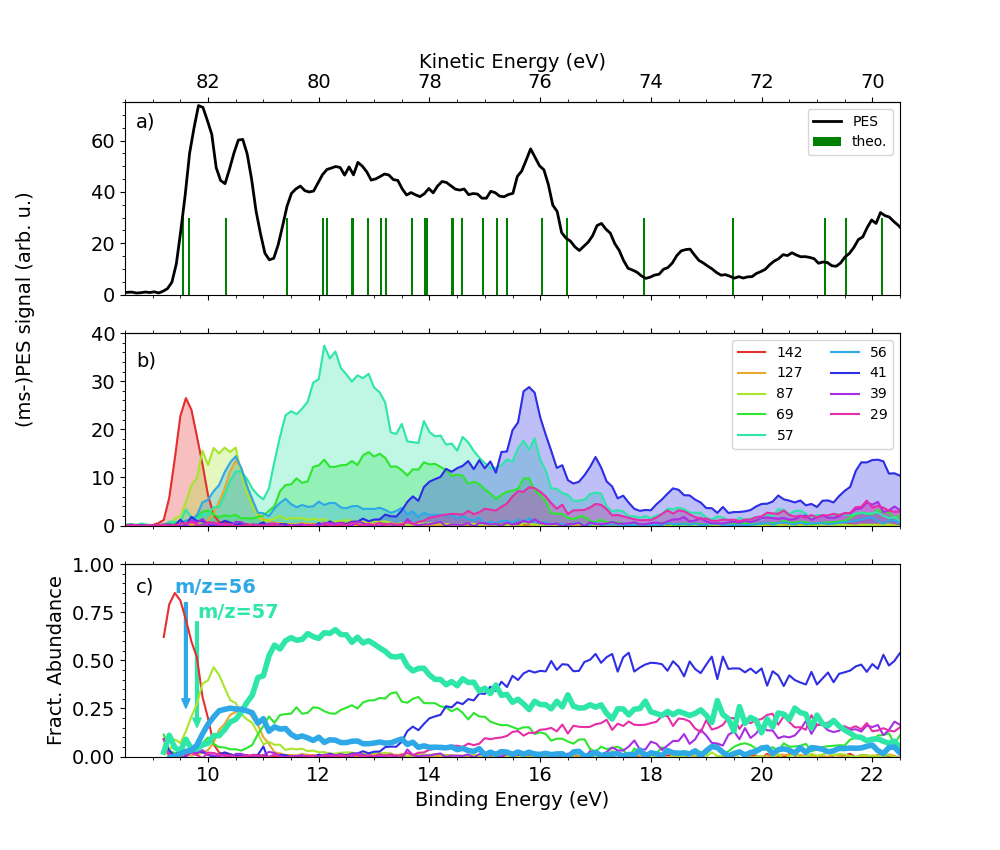}
\end{tabular}
\end{center}
\caption{\label{fig:pes} Photoelectron, PEPICO spectra, and branching ratios for the photoionization of \textit{tert}-butyl methacrylate at 92 eV. (a) Total photoelectron spectrum recorded at 92 eV photon energy (black line). The green bars show computed G$_0$W$_0$ quasi particle orbital energies. (b) Mass-selected photoelectron spectrum for the nine most-abundant ions generated upon 92 eV photoionization. (c) Relative intensities of the electron-ion coincidence signal as a function of binding energy.}
\end{figure}

PEPICO measurements can provide clarification on the question if EUV photoionization directly induces a deprotection reaction in a photoresist model system, since it provides direct information on state-selected dissociation pathways in the cation. The binding energy at which a fragment ion signal starts appearing depends on the underlying fragmentation mechanism and comparison of this so-called appearance energy to quantum-chemical calculations enables the assignment of a peak in the mass spectrum to the chemical structure of the fragment ion. Fig. \ref{fig:pes}a shows the total photoelectron spectrum, i.e., integrated over all fragment ions. The distribution of the kinetic energies of the photoelectrons is broad and, as expected, the majority of primary photoelectrons have a high kinetic energy between 83 and 75~eV for photoionization at 92~eV. With increasing chain length, the energy of the highest occupied molecular orbital with respect to the vacuum level generally increases, leading to lower gas phase ionization energies.\cite{McCormick2013,Larsen2016} However, a net effect in the opposite direction occurs when comparing gas phase to solid phase photoemission spectra due to the instrument work-function, long range bonding, and referencing to the Fermi level instead of the vacuum level for solid and gas phase, respectively. This explains why the kinetic energy distribution of the primary electrons in gas phase photoemission spectra from monomers  and from solid phase polymer thin films is similar.\cite{Ma2019,Zhang2021,Kostko2022} The slope of the onset of the photoelectron signal is rather shallow and no vibrational structure is resolved. This is an indication that the geometry of the parent molecule changes significantly upon photoionization. Density functional theory on the pbe0/def2-tzvp level indeed confirms that the structure of the neutral molecule and the cationic ground state are quite different. The most pronounced change is predicted to occur in the dihedral angle O2-C4-C2-C3. While both double bonds nearly lie in one plane for the neutral molecule (dihedral angle 0.6$^{\circ}$), the dihedral angle is 113.6$^{\circ}$ in the computed structure for the cation. This can be interpreted such that by ionization into the ground state electron density from the conjugated $\pi$-bonds is removed. The observed structure of the photoelectron spectrum originates from ionization out of different molecular orbitals. Quasi-particle energies of these molecular orbitals can be predicted using the G$_0$W$_0$ method using the hybrid pbe0 functional as a starting point. A good qualitative agreement between the experimental spectrum and the computed orbital energies is obtained. From that comparison it becomes clear that the first peak at 9.75 eV contains contributions from the ground state and the first excited state of the cation, since the HOMO and HOMO-1 orbital are separated by only 75 meV in energy. At higher binding energies the agreement between experiment and theory is worse. Here, inner-valence shell ionization occurs for which the $GW$ method is not valid anymore because it only accounts for uncorrelated electronic states .

In panel (b) of Fig. \ref{fig:pes} the total photoelectron spectrum is de-convoluted into the different mass-selected PEPICO spectra. Unsurprisingly, the photoelectrons at the onset of the total electron signal at around 9.2~eV binding energy are recorded in coincidence with undissociated parent ions since the internal energy of the molecular ion is not sufficient to break a chemical bond. However, the relative intensity of the parent ion plotted in Fig. \ref{fig:pes}c does not start out at one because of the low appearance energies of the first daughter ions and the relatively low experimental energy resolution of about 500~meV. In the binding energy region up to 15~eV the fragment m/z=57~amu is dominant before it gets overtaken by m/z=41~amu. As discussed above, the focus of this paper is to answer the question of the origin of m/z=56. If one compares the mass-selected photoelectron spectrum and the fractional abundances of m/z=56 and m/z=57, it becomes evident that the lighter mass m/z=56~amu has a slightly lower appearance energy (9.6$\pm$0.2~eV) than the m/z=57~amu fragment (9.8$\pm$0.5~eV) as listed in \ref{tab:table_branching}. This is already an indication that m/z=56 is not a sequential channel of m/z=57 undergoing a hydrogen loss, but points indeed towards a direct deprotection mechanism by EUV photoionization. In order to further confirm this hypothesis, quantum-chemical computations of the fragmentation pathways of those two channels were performed.

\begin{figure} [ht]
\begin{center}
\begin{tabular}{c} 
\includegraphics[width=15cm]{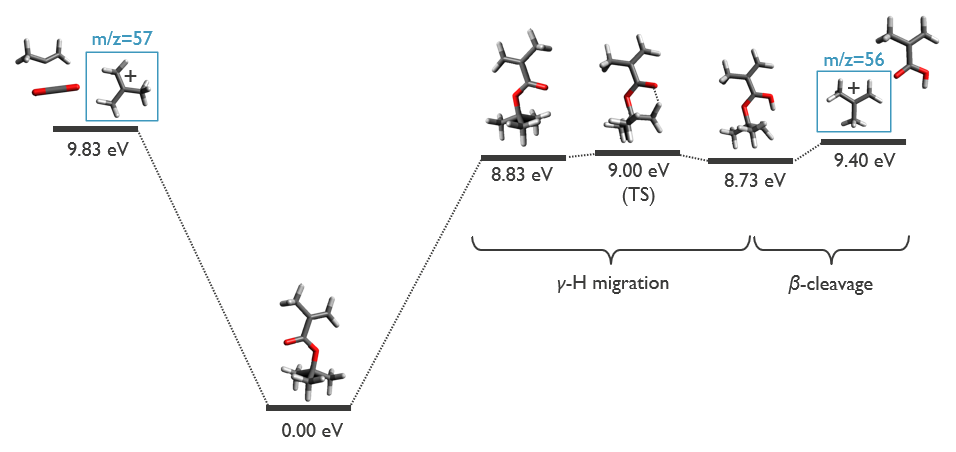}
\end{tabular}
\end{center}
\caption{\label{fig:mechanism} Reaction pathways for dissociative photoionization of \textit{tert}-butyl methacrylate leading to the fragment ions m/z~=~57 (left) and m/z~=~56 (right) computed on the pbe0/def2-tzvp level of theory. }
\end{figure}

A relaxed coordinate scan of the O1-C5 bond in the cation on the pbe0/def2-tzvp level of theory shows that the dissociation leading to the \textit{tert}-butyl cation with m/z~=~57 amu takes place without a reverse barrier as shown on the left of Fig. \ref{fig:mechanism}. Interestingly, the neutral counterpart fragment is not stable and the C2-C4 bond breaks as well leading to the methylvinyl radical H$_3$CCCH$_2$ and CO$_2$. This finding is consistent with EUV outgassing experiments on a model CAR containing a co-polymer with TBMA monomer units, in which the \textit{tert}-butyl and CO$_2$ signal was observed at a constant ratio for different EUV doses.\cite{Pollentier2018} The present study shows that dissociative photoionization can be one of the origins for this unwanted degradation of the polymer hindering a solubility switch. The computations predict an appearance energy of 9.83~eV for the \textit{tert}-butyl cation relative to the neutral \textit{tert}-butyl methacrylate parent molecule, which is in agreement with the observed onset of the PEPICO signal of m/z=57~amu.

Cleaving a C-H bond in the \textit{tert}-butyl cation would lead to isobutene ions with m/z=56~amu and a neutral hydrogen atom. The computations predict an activation energy of 3.67~eV for this sequential fragmentation mechanism, which would lead to an appearance energy for m/z=56~amu of 13.6~eV. This clearly contradicts the experimental observation that m/z=56 and 57 appear at similar binding energies. An alternative mechanism must thus lead to the formation of m/z=56 fragment ions. As can be seen on the right half of Fig. \ref{fig:mechanism}, the migration of a hydrogen atom from the \textit{tert}-butyl group to the carbonyl oxygen has a very low-lying transition state and leads to an energetically favored intermediate state. Breaking of the O1-C5 in the latter then leads to the isobutene cation with m/z=56~amu and neutral methacrylic acid. This mechanism is actually quite common in the dissociation of organic molecules containing a keto group and is usually referred to as McLafferty rearrangement\cite{McLafferty1959} in mass spectrometry and Type II Norrish reaction in photochemistry.\cite{NORRISH1936}
Since none of the intermediates and transition states lie higher in energy than the products, the appearance energy for isobutene cations formed through this mechanism is predicted to be 9.40~eV, i.e., 400~meV below the appearance energy of \textit{tert}-butyl cations. These computations are thus in good agreement with the experiment, where the onset of m/z=56~amu ions is observed at a slightly lower binding energy than the onset of m/z=57~amu ions and confirms the hypothesis above that dissociative photoionization of TBMA can directly deprotect the ester and lead to methacrylic acid. 

\section{DISCUSSION}
\label{sec:discussion}

It can be safely assumed that the observed dissociative photoionization channels also play a role in thin films of a CAR containing TBMA monomer units because the electronic structure is very similar. However, in a thin film recombination of electrons and ions can occur and will thus decrease the final yield of fragmentation by photoionization. The mean free path of primary electrons in photoresists is in the range of 0.5-1~nm,\cite{Seah1979} i.e., comparable to the size of a monomer unit. Electron-ion recombination is thus possible, but not expected to be dominant. Since the majority of photon-molecule/material interaction do not lead to the desired solubility switch, dissociative photoionization is likely one of the sources for the formation of unwanted side products that lead to a decrease of the patterning performance. As obvious from Tab. \ref{tab:table_branching} many of the neutral fragments from dissociative photoionization are reactive radicals that are prone to react and can possibly lead to side reactions such as cross-linking in a polymer.

Thanks to this isolated view on the first reaction step in an EUV exposure, the photoionization, one can develop new strategies on how to improve the patterning performance of photoresists. On the one hand, side reactions, which lead to fragmentation pathways that hinder a solubility switch and lead to products that potentially boost cross-linking, should be suppressed. This is difficult to realize because dissociative photoionization will always be dominant compared to non-dissociative photoionization due to the high energy of EUV photons. On the other hand, one could try to steer the fragmentation pattern towards a channel which directly induces a solubility switch, such as the one leading to m/z=56~amu in the present study. As one can see in the PEPICO spectra in Fig. \ref{fig:pes}, the m/z=56~amu channel is significant almost only in the binding energy region between 9.7 and 11.0~eV. It is thus primarily associated to a single electronic state in the cation - ionization out of the third highest lying occupied molecular orbital (HOMO-2). By modifying the electronic structure of the molecule, e.g., through variation of the substituents one could try to increase the population of this electronic state and thus increase the relative importance of the dissociation channel leading to a direct solubility switch. In this context, ultrafast time-resolved experiments on the femto- to picosecond time scale could help visualize the dynamics that lead to the population of the metastable state associated to this particular fragmentation channel. The photoelectron spectrum depicts a snapshot at the instant of photoionization, while the mass spectrum shows the result after electronic and nuclear motion have lead to fragmentation. PEPICO therefore frames what is happening on ultrafast time scales in photoresists. Gaining more insights into the time between ionization and fragmentation would be an important milestone necessary to finally win control over dissociative photoionization.

While this study focuses on a monomer unit widely used in CARs, dissociative photoionization is supposedly even more relevant in non-CAR materials, in particular ones which have metal atoms with high absorption cross sections in the EUV incorporated. For non-CARs the solubility switch is usually not controlled by a catalytic reaction, but bond scissions and cross-linking. Studying the dissociative photoionization of metal-oxide resists or other non-CAR resist classes will have an important impact to better understand the exposure chemistry in state-of-the-art EUV photoresists.

As discussed in the introduction, taking advantage of further complementary techniques like electron impact ionization or EUV outgassing experiments will round off this study on dissociative photoionization and enable to draw a more complete picture of the EUV resist chemistry to consider also the chemistry induced by low energy secondary electrons and diffusion processes.

\section{CONCLUSIONS}
\label{sec:conclusions}

The dissociative photoionization of \textit{tert}-butyl methacrylate, a molecule that is found as a monomer unit in common CAR systems, was studied in the gas phase using EUV synchrotron radiation and photoelectron photoion coincidence detection. This experiment enabled an isolated view on the interaction of EUV photons with matter, which is not possible in the solid state. It was found that more than 96~\% of the molecules interacting with EUV photons fragment undergoing dissociative photoionization. While the most important channel leads to the formation of \textit{tert}-butyl cations, neutral 2-methylvinyl radicals and carbon dioxide, a fragmentation path for direct deprotection of the ester leading to isobutene cations and methacrylic acid was identified thanks to the detailed insights gained in PEPICO experiments combined with quantum chemical computations. These findings show that dissociative photoionization can on the one hand be one of the sources for unwanted side reactions lowering the performance of a photoresist, but on the other hand can also provide a knob for better control of the chemistry induced by the interaction of EUV photons with matter. 

\acknowledgments 
\label{sec:acknowledgement}
MGG acknowledges financial support The Research Foundation - Flanders (FWO) for a travel grant. MG acknowledges funding by the Deutsche Forschungsgemeinschaft, contract F1575/13-2.

\bibliography{TBMA_library} 
\bibliographystyle{spiebib} 

\end{document}